\documentclass[11pt]{article}
\usepackage{graphicx} 
\usepackage{epsfig} 
\usepackage{dcolumn}
\usepackage{bm}
\usepackage{amssymb}
\usepackage{amsfonts}
\usepackage{amsthm}
\usepackage{latexsym} 
\usepackage{cite} 

\begin{document}

\title{Nuclear energy functional with a surface-peaked effective mass:
       Global properties}

\author{A. F. Fantina$^{1,2}$, J. Margueron$^1$, P. Donati$^2$ and P. M. Pizzochero$^2$\\
$^1$Institut de Physique Nucl\'eaire, Universit\'e Paris-Sud, IN2P3-CNRS,\\ 
F-91406 Orsay Cedex, France,\\
$^2$Dipartimento di Fisica, Universit\`a degli Studi di Milano, \\
and Istituto Nazionale di Fisica Nucleare, Sezione di Milano, \\
Via Celoria 16, 20133 Milano, Italy}

\maketitle

\begin{abstract}
A correction to the nuclear functional is proposed in order to improve the density of 
states around the Fermi surface. 
The induced effect of this correction is to produce a surface-peaked effective mass, 
whose mean value can be tuned to get closer to 1 for the states close to the Fermi energy. 
In this work we study the effect of the correction term on global properties of nuclei such 
as the density of states in $^{40}$Ca and $^{208}$Pb, pairing
and specific heat at low temperature in $^{120}$Sn. 
In the latter application, an explicit temperature-dependent form of the correction term is 
employed and it is shown that the critical temperature is reduced by 40-60~keV.
\end{abstract}



In finite nuclei, the single particle states around the Fermi
energy are known to be strongly affected by the dynamical
particle-hole correlations~\cite{mahaux}. The energy of the single
particle states is modified and thus the level density around the
Fermi energy is changed, which in turn has an impact on low-energy
properties such as pairing
correlations~\cite{dean2003,sagawa2008}, collective
modes~\cite{harakeh2001} as well as on temperature-related
properties such as the entropy, the critical temperature, the
specific heat~\cite{chamel2009}. In nuclear astrophysics, a good
description of the density of states around the Fermi energy
(which is related to the nucleon effective mass) turned out to be
relevant also for the energetics
 of  core-collapse supernovae~\cite{donati1994,fantina2008}. \\

Despite the important role of the density of states in
self-consistent mean field theories, most of these models such as
those based on Skyrme~\cite{skyrme1956}, Gogny~\cite{gogny1975}
and M3Y interactions~\cite{nakada2003} or on the relativistic
approaches like RMF~\cite{serot1997} or RHF~\cite{long2006}, have
a density of states around the Fermi energy that is too low.
Already in the 1960s, G.E. Brown \textit{et al}. suggested that the
effective mass $m^*/m$ should be close to 1 around the Fermi
energy in order to improve
the level density~\cite{brown1963}. 
It could indeed be shown that the first order expansion of the
energy-dependent self-energy around the Fermi
surface~\cite{migdal1967,ma1983} produces an effective mass which
is the product of two different terms, the $k$-mass $m_k$ and the
$\omega$-mass $m_\omega$~\cite{mahaux}. The $\omega$-mass is
related to the energy dependent part of the self-energy,
dynamically generated by the coupling of the particles to the
core-vibrations~\cite{bertsch1968,bernard1979,brown1979,bortignon1987}.
These dynamical correlations, which lead to an increase of the
effective mass at the surface, are 
implemented beyond the
mean field. Indeed, 
the effective mean field theories have
an average effective mass $m^*/m$,
the $k$-mass, 
around 0.6-0.8.

The different microscopic calculations of the particle-vibration
coupling (PVC) have been performed at the first order in
perturbation, due to their heavy computational features
\cite{bertsch1968,bernard1979,brown1979,bortignon1987,giai1983,ring2009,yoshida2009}.
However, the induced effects of the PVC on the mean field itself
is non-negligible. Since the effective mass is enhanced at the
surface of nuclei~\cite{giai1983}, it impacts the single particle
states and the pairing correlations which, in turn, modify the PVC
and the effective mass. There is then a self-consistent relation
between the properties of the single particle states around the
Fermi energy and the PVC.

In this paper, we propose an effective short-cut for treating the
effects of the PVC directly in the Energy Density Functional (EDF)
approach . The coupling of the collective modes to the single
particle motion induces a dynamical type of correlation that in
principle could not be easily implemented in an effective nuclear
interaction or in a nuclear EDF.
However, the core-vibration is mainly located at the surface of
the nuclei which makes the implementation in energy-density
functionals easier. 
A parametrization of the
$\omega$-mass as a gradient of the nuclear profile has shown to
give good results within a nuclear shell model~\cite{ma1983}. In
addition, the first order expansion of the self-energy near the
Fermi energy induces a renormalization of the single particle
Green function as well as a correction to the mean field, which
almost compensate the effective component of the equivalent
potential~\cite{ma1983}.
In the present approach, the surface-peaked effective mass is
included through a correction term in the Skyrme energy density
functional. This correction is energy independent and designed in
such a way to have a moderate effect on the mean field.

The paper is organized as follows: in Sec.~\ref{sec2} we will
describe the theoretical framework. In Sec.~\ref{sec3} we will
discuss the application to spherical nuclei, in particular, to
${}^{40}\mathrm{Ca}$ and ${}^{208}\mathrm{Pb}$. In Sec.~\ref{sec4}
we discuss the superfluid properties in connection with the
surface-peaked effective mass, both at $T=0$ and at finite
temperature. Finally, in Sec.~\ref{sec6} we will give our
conclusions and outlooks.

\section{\label{sec2} Nuclear energy density functional}

The energy density functional derived from the Skyrme interaction will be 
considered as follows.
The standard Skyrme energy-density $\mathcal{H}(\mathbf{r})$ is expressed in terms of a 
kinetic term $\mathcal{K}(\mathbf{r})$ and an interaction term written as~\cite{bender2003}
\begin{equation}
\mathcal{H}(\mathbf{r}) = \mathcal{K}(\mathbf{r}) + \sum_{T=0,1} \mathcal{H}_T(\mathbf{r}) \ ,
\end{equation}
with:
\begin{eqnarray}
\mathcal{K}(\mathbf{r}) &=& \frac{\hbar^2}{2m} \tau(\mathbf{r}) \;, \\
\mathcal{H}_T(\mathbf{r}) &=& C^\rho_T \rho^2_T(\mathbf{r}) + C^{\nabla^2 \rho}_T \rho_T(\mathbf{r}) \nabla^2 \rho_T(\mathbf{r}) 
+ C^\tau_T \rho_T(\mathbf{r}) \tau_T(\mathbf{r}) \nonumber \\
&+& C^J_T \mathbb{J}^2_T(\mathbf{r}) + C^{\nabla J}_T \rho_T(\mathbf{r}) \nabla \cdot \mathbf{J}_T(\mathbf{r})  \;,
\label{eq:ht}
\end{eqnarray}
where the kinetic density $\tau(\mathbf{r})$, the density $\rho_T(\mathbf{r})$, the spin-current
$\mathbb{J}^2_T(\mathbf{r})$ and the current $\mathbf{J}(\mathbf{r})$ as well as the relation between
the Skyrme parameters and the coefficients in Eq.~(\ref{eq:ht}) are defined in Ref.~\cite{bender2003}.
We have kept only the time-even component for the application considered in this paper.
Notice the difference between the integrated energy, $H$, and the energy density functional, $\mathcal{H}$.
These two quantities are indeed related as: $H=\int \! d\mathbf{r} \, \mathcal{H}(\mathbf{r})$.

In the following, we explore the impact of adding to the functional an iso-scalar correction of the form:
\begin{eqnarray}
\mathcal{H}_0^\mathrm{corr}(\mathbf{r}) = C^{\tau (\nabla \rho)^2}_0 \tau(\mathbf{r}) \left(\nabla \rho(\mathbf{r})\right)^2 
+C^{\rho^2 (\nabla \rho)^2}_0 \rho(\mathbf{r})^2 \left(\nabla \rho(\mathbf{r})\right)^2 \ ,
\label{eq:hcorr}
\end{eqnarray}
where the first term
have been first introduced in Ref.~\cite{zalewski2010,zalewski2010_bis} and
induces a surface-peaked effective mass while the second term is introduced 
to moderate the effect of the first one in the mean field.

The effective mass is obtained from the functional 
derivative
%
of the energy $H$ and is expressed as 
($q$ runs over neutrons and protons: $q = n,p$):
\begin{equation}
\frac{\hbar^2}{2 m_q^*(\mathbf{r})} \equiv \frac{\delta H}{\delta \tau_q} = \frac{\hbar^2}{2 m} + C^\tau_q \rho_q(\mathbf{r}) 
+ C^{\tau (\nabla \rho)^2}_0 (\nabla \rho(\mathbf{r}))^2 \ ,
\end{equation}
where the coefficient $C^\tau_q=(C^\tau_0 \pm C^\tau_1)/2$ (with $+$ for $q=n$ and $-$ for $q=p$),
and the mean field reads:
\begin{equation}
U_q(\mathbf{r}) = U_q^\mathrm{Sky}(\mathbf{r}) + U^\mathrm{corr}(\mathbf{r}) \ , 
\end{equation}
where $U_q^\mathrm{Sky}(\mathbf{r})$ is the mean field deduced from the Skyrme interaction~\cite{bender2003}
and $U^\mathrm{corr}(\mathbf{r})$ is the correction term induced by Eq.~(\ref{eq:hcorr}) which is defined as:
\begin{eqnarray}
U^\mathrm{corr}(\mathbf{r}) &=& -2 C^{\tau (\nabla \rho)^2}_0 \Big( \tau(\mathbf{r}) \nabla^2 \rho(\mathbf{r}) 
+ \nabla \tau(\mathbf{r}) \nabla \rho(\mathbf{r}) \Big) \nonumber \\
&-& 2 C^{\rho^2 (\nabla \rho)^2}_0 \Big( \rho(\mathbf{r}) (\nabla \rho(\mathbf{r}))^2 
+ \rho(\mathbf{r})^2 \nabla^2 \rho(\mathbf{r}) \Big) \; .
\label{eq:ucorr}
\end{eqnarray}

\bigskip
From the variation of the total energy $H$ with respect to the ground state density 
matrix
 we obtain the following set of self-consistent Kohn-Sham equations:
\begin{eqnarray}
 \left[ - \nabla \cdot \frac{\hbar^2}{2 m_q^*(\mathbf{r})} \nabla + U_q(\mathbf{r}) - i W_q(\mathbf{r}) \cdot (\nabla \times \sigma) \right] \Phi_{\lambda,q}(\mathbf{r})  = \epsilon_{\lambda,q}\ \Phi_{\lambda,q}(\mathbf{r}) \ ,
\label{eq:ks}
\end{eqnarray}
where $\lambda$ runs over neutron and proton orbitals.

For spherical nuclei, the single particle wave functions $\Phi_{\lambda,q}$ can be factorized 
into a radial part and an angular part as (cf. Eq.~(26) of Ref.~\cite{vautherin1972}):
\begin{equation}
\Phi_{\lambda,q}(\mathbf{r},\sigma,\tau) = \frac{\phi_{\lambda,q}(r)}{r} \ \mathcal{Y}_{l,j,m}(\theta,\phi,\sigma)\ \chi_q(\tau) \ ,
\end{equation}
where $\sigma$ is the spin and $\tau$ the isospin, and the radial wave function satisfies the following equation:
\begin{eqnarray}
&&\left[-\frac{\hbar^2}{2 m} \frac{\mathrm{d}^2}{\mathrm{d}r^2} + \frac{\hbar^2}{2 m} \frac{l (l+1)}{r^2} + V^\mathrm{eq}_q(r,\epsilon) \right] \psi_{\lambda,q}(r) = \epsilon_{\lambda,q} \psi_{\lambda,q}(r) \ ,
\label{eq:radial}
\end{eqnarray}
where $\phi_{\lambda,q}$ differs from the solution of Eq.~(\ref{eq:radial}), $\psi_{\lambda,q}(r)$, by a 
normalization factor, $\phi_{\lambda,q}(r) = (m^*_q(r)/m)^{1/2} \psi_{\lambda,q}(r)$. 
The equivalent potential $V_q^\mathrm{eq}$ is usually introduced for practical reasons, 
\begin{eqnarray}
V_q^\mathrm{eq} (r,\epsilon) &=&  \frac{m^*_q(r)}{m} \Big[ V_q(r) + U_q^{so}(r) \langle \mathbf{l} \cdot \mathbf{\sigma} \rangle + \delta_{q,p} V_{Coul}(r) \Big]  +  \left[1 - \frac{m^*_q(r)}{m} \right] \epsilon_{\lambda,q} \ , \nonumber\\
\end{eqnarray}
where $U_q^{so}(r)$ is the spin-orbit potential~\cite{bender2003}, $V_{Coul}$ is the Coulomb potential, and
\begin{eqnarray}
V_q(r) &=& U_q^\mathrm{Sky}(r) + U^\mathrm{corr}(r) + U_q^\mathrm{eff}(r) \; , \label{eq:vq} \\
U_q^\mathrm{eff} &=& - \frac{1}{4} \frac{2 m^*_q(r)}{\hbar^2} \left( \frac{\hbar^2}{2 m^*_q(r)} \right)'^2 + \frac{1}{2} \left( \frac{\hbar^2}{2 m^*_q(r)} \right)'' 
           + \left( \frac{\hbar^2}{2 m^*_q(r)} \right)' \frac{1}{r} \label{eq:ueff} \ .
\end{eqnarray}

\bigskip
In the original work of Ma and Wambach~\cite{ma1983}, the term $U^\mathrm{corr}(r)$ was derived directly from
the Green's function with energy dependent self-energies while, in our approach, $U^\mathrm{corr}(r)$ 
is derived from the new term~(\ref{eq:hcorr}) in the EDF.
A one-to-one correspondence between EDF and the Green's function approach is not possible.
However, since the terms $U^\mathrm{corr}(r)$ and $U_q^\mathrm{eff}(r)$ compensate each other in 
the Green's function approach~\cite{ma1983}, we want to reproduce the same behavior in the EDF.
We obtain approximate compensation 
%
between $U^\mathrm{corr}(r)$ and $U_q^\mathrm{eff}(r)$ by 
imposing the following relation between the new coefficients: 
\begin{equation}
C_0^{\rho^2 (\nabla \rho)^2} = 12\ \mathrm{fm}\ C_0^{\tau (\nabla \rho)^2} \ .
\end{equation}
We have investigated the sensitivity of the results to the value of the proportionality constant and we have checked that the reasonable values lie in the range 10-20 fm, after which the compensation is no longer efficient.

The effects of the correction terms in (\ref{eq:hcorr}) will be analyzed in the next section.
Notice that
%
a surface-peaked effective mass could also be 
obtained from a modified Skyrme interaction.
This different approach
potentially leads to an improved 
agreement with experimental single particle energies~\cite{farine2001}.
Since the new term explored in Ref.~\cite{farine2001} is simultaneously momentum and
density dependent, the functional obtained is quite different from Eq.~(\ref{eq:hcorr}): 
the number of terms is much larger and the correction to the effective mass is a 
polynomial in the density.
It would be interesting to carry out a more detailed comparison of these two different 
approaches in a future study.

\section{\label{sec3} Mean field properties}

In the following, we study the influence of the correction introduced by the new term 
$\mathcal{H}_0^\mathrm{corr}$ for two representative nuclei: $^{40}$Ca and $^{208}$Pb, 
using BSk14 interaction~\cite{goriely2007}, which is adjusted to a large number of 
nuclei (2149).
The effective mass in symmetric matter at saturation density is 0.8$m$ and the isospin
splitting of the effective mass in asymmetric matter qualitatively agrees with the 
expected behavior deduced from microscopic Brueckner-Hartree-Fock 
theory~\cite{vandalen2005}.
In the following, we study the effect of the correction term on top of BSk14
interaction without refitting the parameters.
The refit is in principle necessary since the correction term impacts the
masses and changes the single particle energies.
In this first exploratory work, we improve the level density
and discuss the effects on other quantities like the pairing, the entropy and the specific heat.
The correction term~(\ref{eq:hcorr}) could however potentially bring a better 
systematics in the comparison to experimental single particle centroids.
This will be studied in a future work.

\begin{figure}[htb]
\begin{center}
\includegraphics[width=12cm]{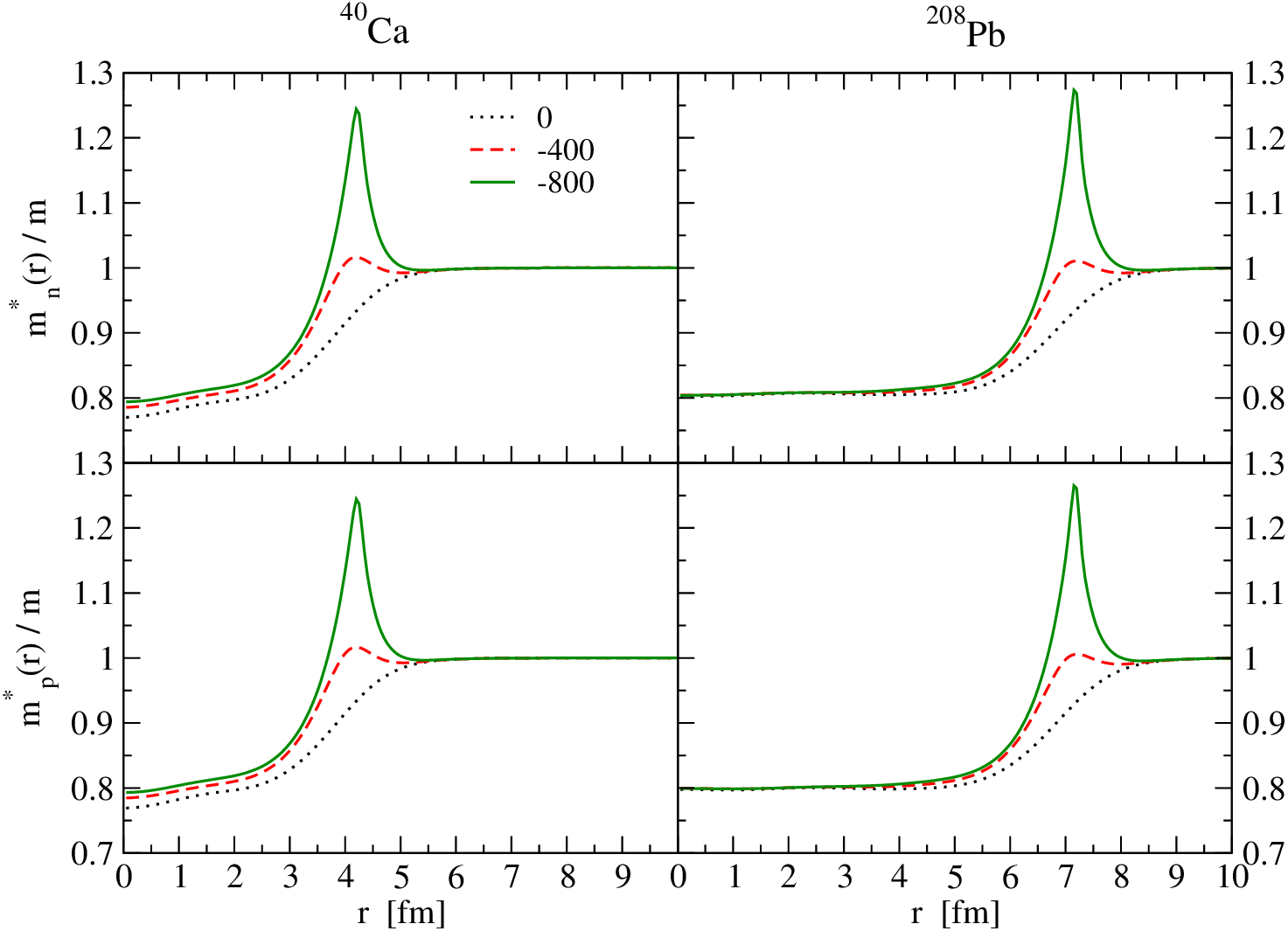}
\end{center}
\caption{$m^*_q / m$ as a function of radial coordinate for ${}^{40}\mathrm{Ca}$ and 
${}^{208}\mathrm{Pb}$, for $C_0^{\tau (\nabla \rho)^2} =$ 0, -400, -800~MeV fm$^{10}$.}
\label{fig:effmass}
\end{figure}

The neutron and proton effective masses are plotted in Fig.~\ref{fig:effmass} for different values of the 
coefficient $C_0^{\tau (\nabla \rho)^2} =$ 0, -400, -800~MeV fm$^{10}$. 
Notice that the case $C_0^{\tau (\nabla \rho)^2} =$ 0 is that of the original Skyrme interaction BSk14. 
Increasing $|C_0^{\tau (\nabla \rho)^2}|$ from 0 to 800~MeV fm$^{10}$, 
we observe an increase of the effective mass $m^*/m$ at the surface, which produces a peak for large values of  
$C_0^{\tau (\nabla \rho)^2}$, while at the center of the nucleus the effective masses get closer to 
the values obtained without the correction term.
Fig.~\ref{fig:effmass} can be compared with Fig.~1 of Ref.~\cite{ma1983}. 
Due to the different surface-peaked functions (here $(\nabla\rho)^2$ instead of a single $\nabla$ dependence in 
Ref.~\cite{ma1983}), the width of the effective mass at the surface is larger in Ref.~\cite{ma1983} than 
in the present work. 
For values of the coefficient $|C_0^{\tau (\nabla \rho)^2}|$ larger than 800~MeV fm$^{10}$, 
the potentials $U^\mathrm{corr}(r)$ and $U_q^\mathrm{eff}(r)$ (Eq.~(\ref{eq:ucorr}),~(\ref{eq:ueff})) 
induce large gradients of the mean field~(\ref{eq:vq}) in a tiny region close to the surface of the 
nuclei, which in turn produce an instability in the HF iterations.

\begin{figure}[htb]
\begin{center}
\includegraphics[width=12cm]{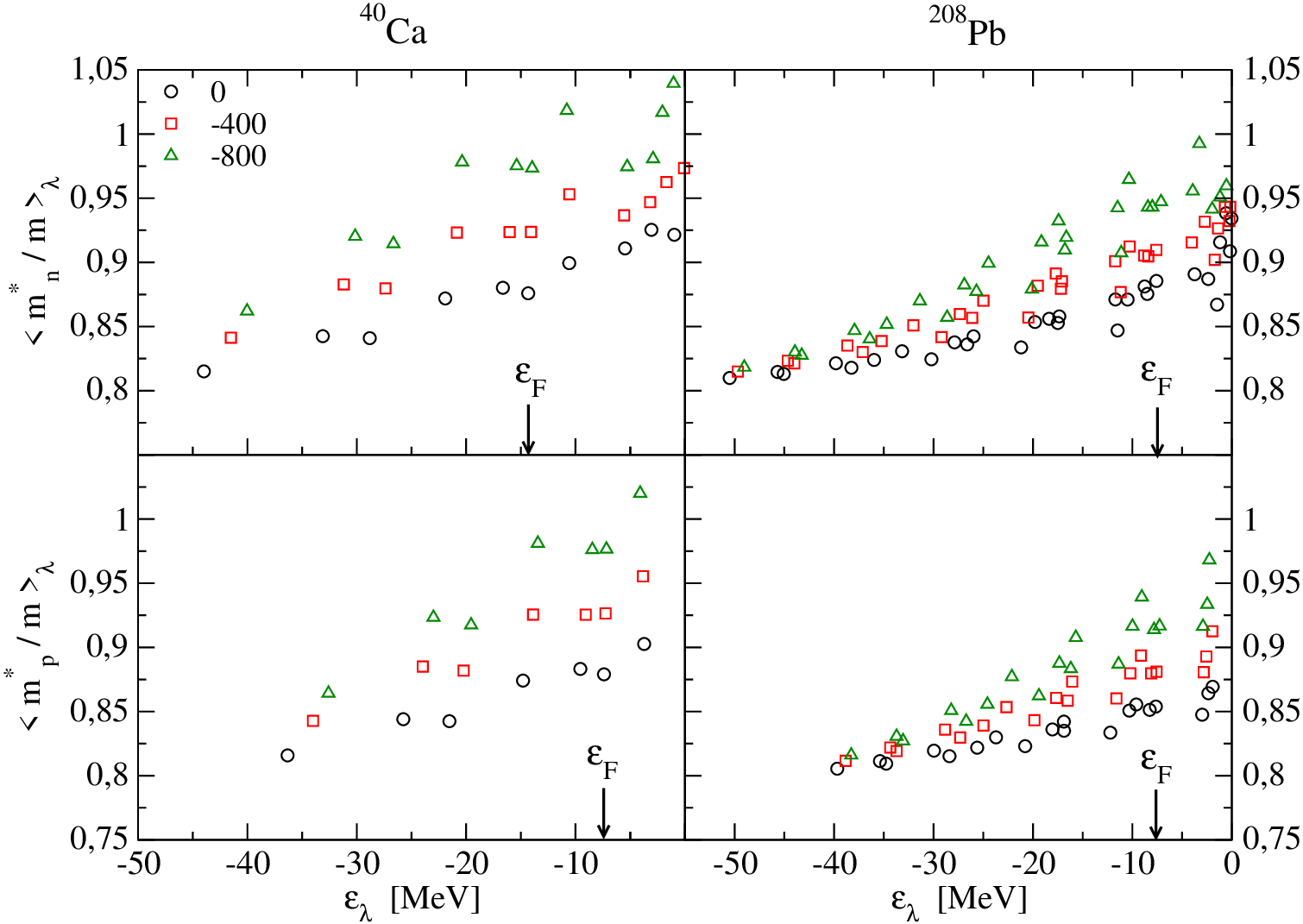}
\end{center}
\caption{Energy dependence of the expectation value of $\langle m^*_q/m\rangle_\lambda$ for bound states 
for $^{40}$Ca and $^{208}$Pb, for $C_0^{\tau (\nabla \rho)^2} =$ 0, -400, -800~MeV fm$^{10}$.
The arrows indicate the position of the Fermi energy for $C_0^{\tau (\nabla \rho)^2} =$ 0~MeV fm$^{10}$
defined as the energy of the last occupied state.
The correction induced by the surface-peaked effective mass produces an increase of the Fermi energy
by 400~keV at most.}
\label{fig:mefav}
\end{figure}

A surface-peaked effective mass has also been deduced from the particle-vibration coupling
within the HF+RPA framework~\cite{giai1983}. 
In such an approach, the effective mass is shown to be peaked not only at the surface of the 
nuclei, but also in a window around the Fermi energy of $\pm$~5~MeV. 
Such an energy dependence could not be implemented straightforwardly in the EDF framework.
We could however evaluate the state-averaged effective mass, $\langle m^*_q/m\rangle_\lambda$, 
defined as:
\begin{equation}
\langle m^*_q/m\rangle_\lambda = \int \! d\mathbf{r} \; \phi^*_{\lambda,q}(\mathbf{r}) \frac{m^*_q(\mathbf{r})}{m} \phi_{\lambda,q}(\mathbf{r}) \; , 
\end{equation} 
where the index $\lambda$ stands for the considered state.
The state-averaged effective masses $\langle m^*_q/m\rangle_\lambda$ are represented in Fig.~\ref{fig:mefav}
as a function of the energy of the bound states, and for different values of the parameter
$C_0^{\tau (\nabla \rho)^2}$.
As the value of the coefficient $|C_0^{\tau (\nabla \rho)^2}|$ gets larger, the state-averaged effective masses 
$\langle m^*_q/m\rangle_\lambda$ approach 1 around the Fermi energy while it 
get closer to the original effective mass
for deeply bound states.
Notice however quantitative differences between Ca and Pb for states around the Fermi energy as well as deeply bound states.
In conclusion, even if we did not introduce an explicit energy dependence of the surface-peaked effective mass, we still find that the expected behavior~\cite{giai1983} of $\langle m^*_q/m\rangle_\lambda$ as a function of energy is qualitatively reproduced.

\begin{figure}[htb]
\begin{center}
\includegraphics[width=12cm]{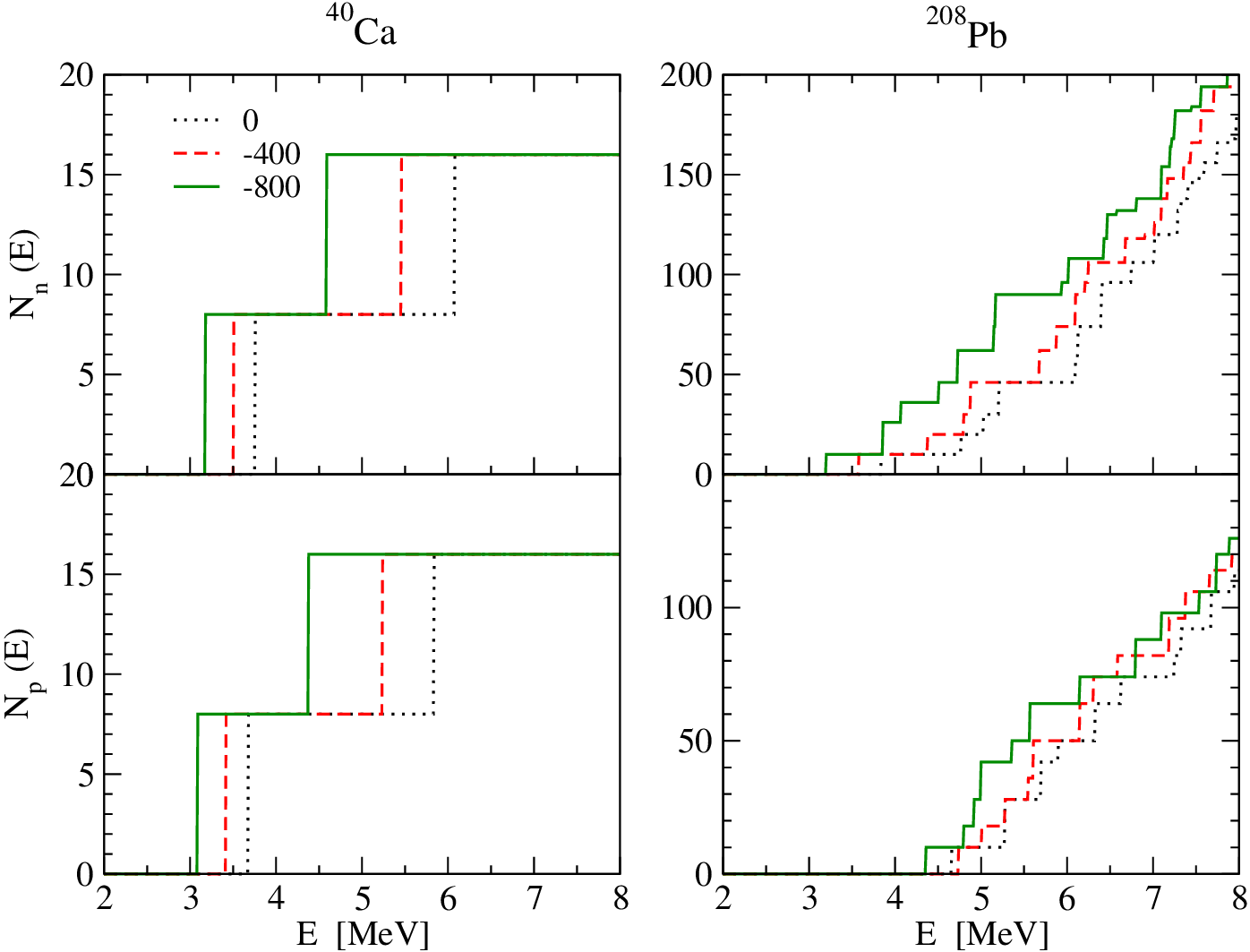}
\end{center}
\caption{Number of states as a function of the excitation energy for $^{40}$Ca and $^{208}$Pb, 
for $C_0^{\tau (\nabla \rho)^2} =$ 0, -400, -800~MeV fm$^{10}$.}
\label{fig:levdens}
\end{figure}

In order to evaluate the impact of the new term on the density of states, we display in Fig.~\ref{fig:levdens} the neutron and proton number of states as a function of the excitation energy, defined as:
\begin{equation}
N(E) = \int_0^E dE'\ g(E') \ ,
\label{eq:ne}
\end{equation}
where g(E) is the density of states,
\begin{equation}
g(E) \equiv \frac{\mathrm{d}N(E)}{\mathrm{d}E} = \sum_{\lambda_1 < F, \lambda_2 > F} \ (2 j_{\lambda_2}+1)\ \delta(E - (\epsilon_{\lambda_2}-\epsilon_{\lambda_1})) \ ,
\label{eq:ge}
\end{equation}
and $\epsilon_{\lambda_1}$ ($\epsilon_{\lambda_2}$) represent the single-particle energies below (above) the Fermi surface.
%
The expected relation between the surface-peaked effective mass and the density of states is clearly shown
in Fig.~\ref{fig:levdens}: the number of states at given excitation energy increases as the coefficient 
$C_0^{\tau (\nabla \rho)^2}$ goes from 0 to -800, meaning that the density of states 
also increases as $|C_0^{\tau (\nabla \rho)^2}|$ gets larger.

\begin{figure}[htb]
\begin{center}
\includegraphics[width=12cm]{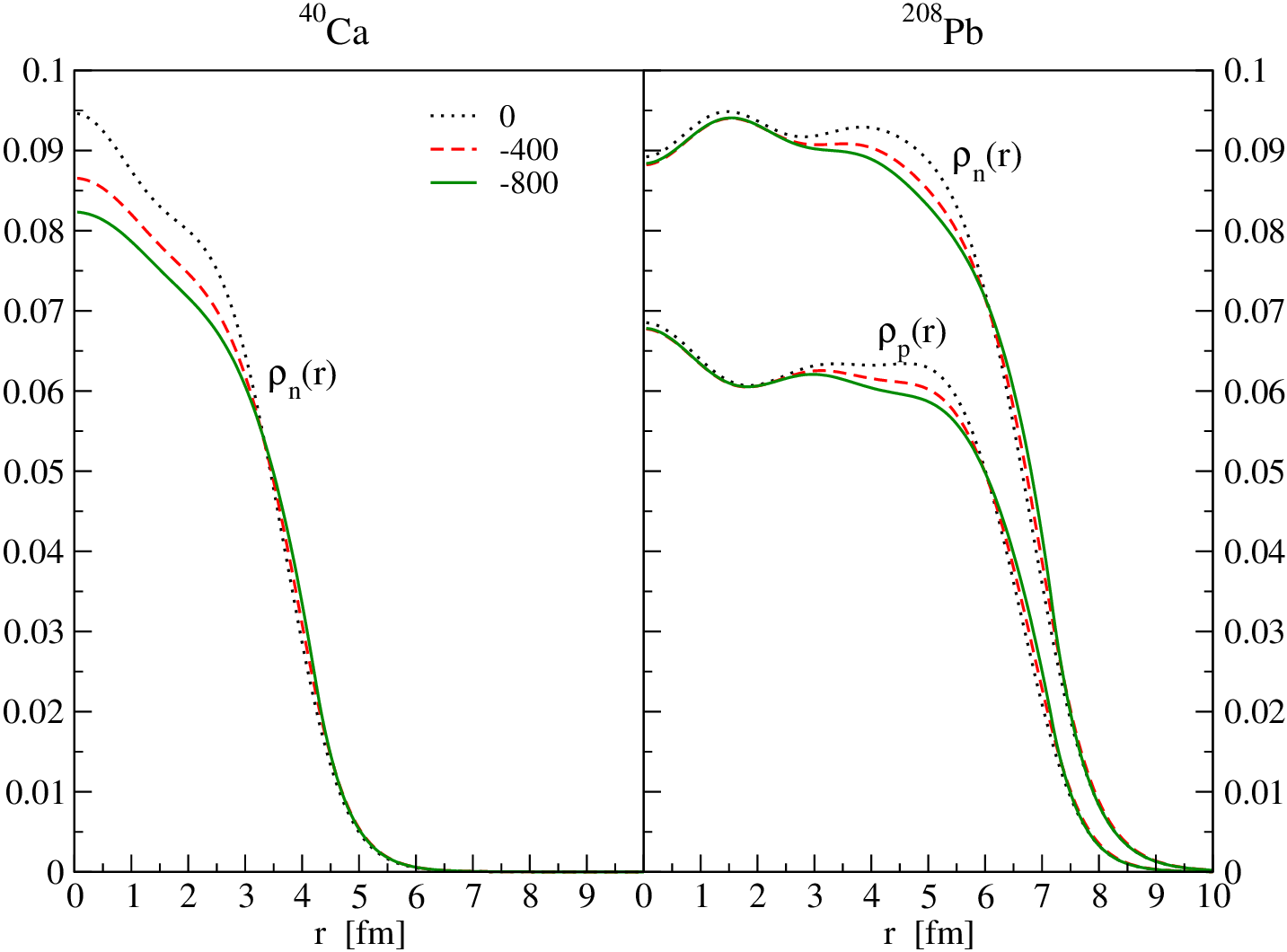}
\end{center}
\caption{Neutron density as a function of radial coordinate for ${}^{40}\mathrm{Ca}$ and neutron 
and proton densities as a function of the radial coordinate for ${}^{208}\mathrm{Pb}$, for 
$C_0^{\tau (\nabla \rho)^2} =$ 0, -400, -800~MeV fm$^{10}$.}
\label{fig:dens}
\end{figure}

Let us now discuss qualitatively the impact of the correction term~(\ref{eq:hcorr}) on global 
properties of nuclei starting with the density profiles.
The neutron and proton densities are shown in Fig.~\ref{fig:dens}.
Since for ${}^{40}\mathrm{Ca}$ the neutron density is very similar to the proton one, we have chosen 
to represent only the neutron one.
Since the number of particles has to be conserved, lower values of the density in the bulk of nuclei for larger values of $|C_0^{\tau (\nabla \rho)^2}|$ are compensated by a slight increase of the size of the nucleus.
These small differences in the density profile influence the charge 
root mean square radius $r_{ch}$ (see Table~\ref{tab:rms}), which slightly increases as the value of the parameter $|C_0^{\tau (\nabla \rho)^2}|$ gets larger.
Moreover, because of the isoscalar nature of the correction~(\ref{eq:hcorr}), neutrons and protons 
are affected in an identical way, as one can see from the constant value of the neutron skin 
radius, $r_{skin}$, given in Table~\ref{tab:rms}.

\begin{figure}[htb]
\begin{center}
\includegraphics[width=12cm]{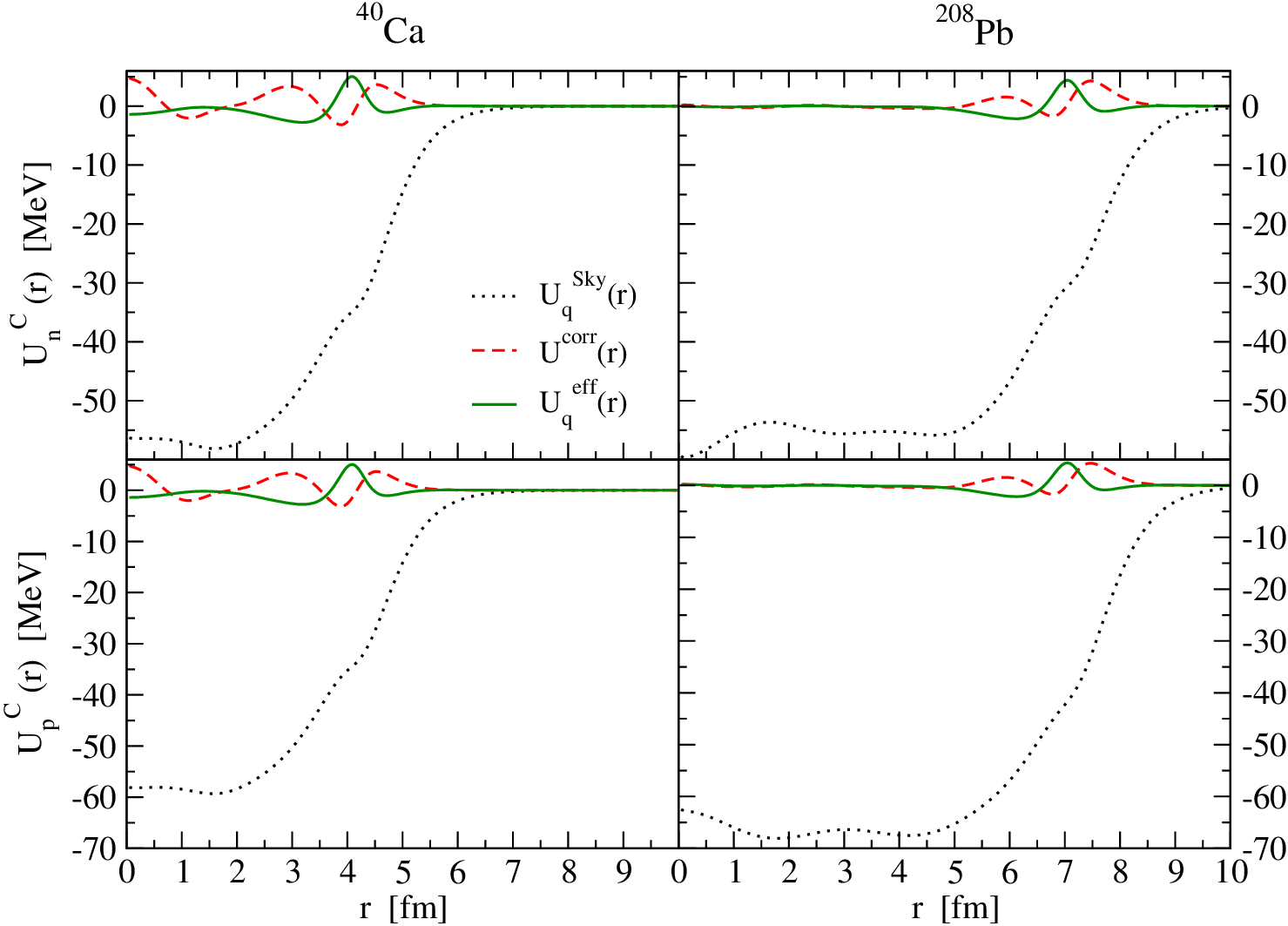}
\end{center}
\caption{Central part of the neutron and proton mean field as a function of radial 
coordinate for $^{40}$Ca and $^{208}$Pb, for $C_0^{\tau (\nabla \rho)^2} = -400$~MeV fm$^{10}$.}
\label{fig:uc}
\end{figure}

Let us now analyze the influence of $\mathcal{H}^\mathrm{corr}_0$ at the level of the mean field $V_q(r)$
defined in Eq.~(\ref{eq:vq}). 
The different components of the central part of the mean field, $U_q^\mathrm{Sky}(r)$, $U_q^\mathrm{eff}(r)$,
and $U^\mathrm{corr}(r)$, see Eqs.~(\ref{eq:ucorr}) and (\ref{eq:ueff}), are represented 
in Fig.~\ref{fig:uc} for the neutrons, the protons, and for $^{40}$Ca and $^{208}$Pb nuclei.
The value of the coefficient is fixed to be $C_0^{\tau (\nabla \rho)^2} = -400$~MeV fm$^{10}$.
As expected, there is a reasonable compensation between $U_q^\mathrm{eff}(r)$ and $U^\mathrm{corr}(r)$.
As a consequence, 
%
the mean field $V_q(r)$ is nearly not affected by the presence of a surface-peaked effective
mass for the values of the coefficient $C_0^{\tau (\nabla \rho)^2}$ chosen in the domain going from 
0 to -800~MeV fm$^{10}$,
except close to the surface where a small change in the slope is observed.
We have then shown that in the EDF framework, the correction term~(\ref{eq:hcorr}) reproduces the result obtained in Ref.~\cite{ma1983}.

\begin{figure}[htb]
\begin{center}
\includegraphics[width=12cm]{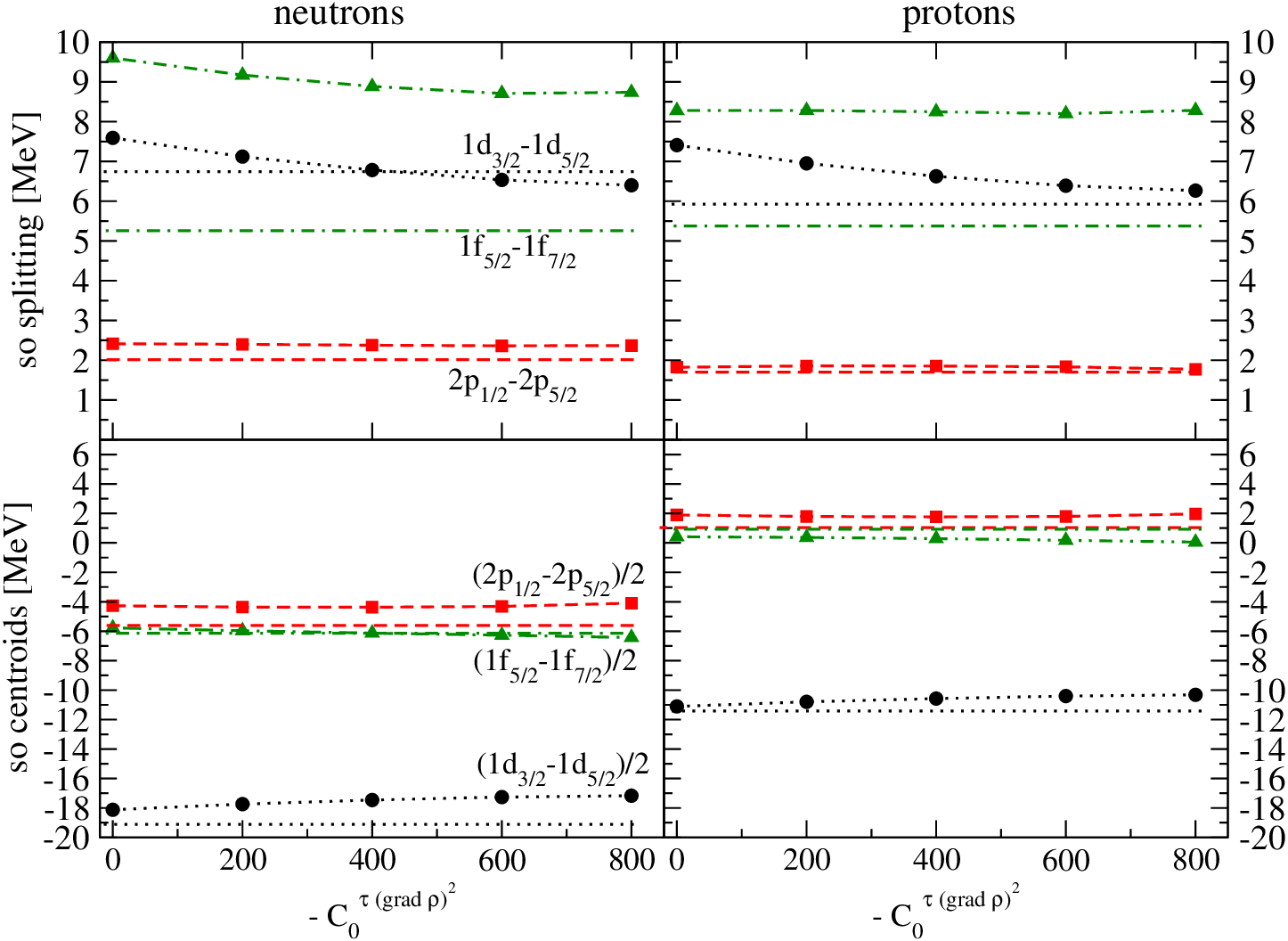}
\end{center}
\caption{Spin-orbit splitting and centroids for $^{40}$Ca as a function of the coefficient 
$C_0^{\tau (\nabla \rho)^2}$. The experimental values are taken from \cite{oros}.}
\label{fig:caso}
\end{figure}

\begin{figure}[htb]
\begin{center}
\includegraphics[width=12cm]{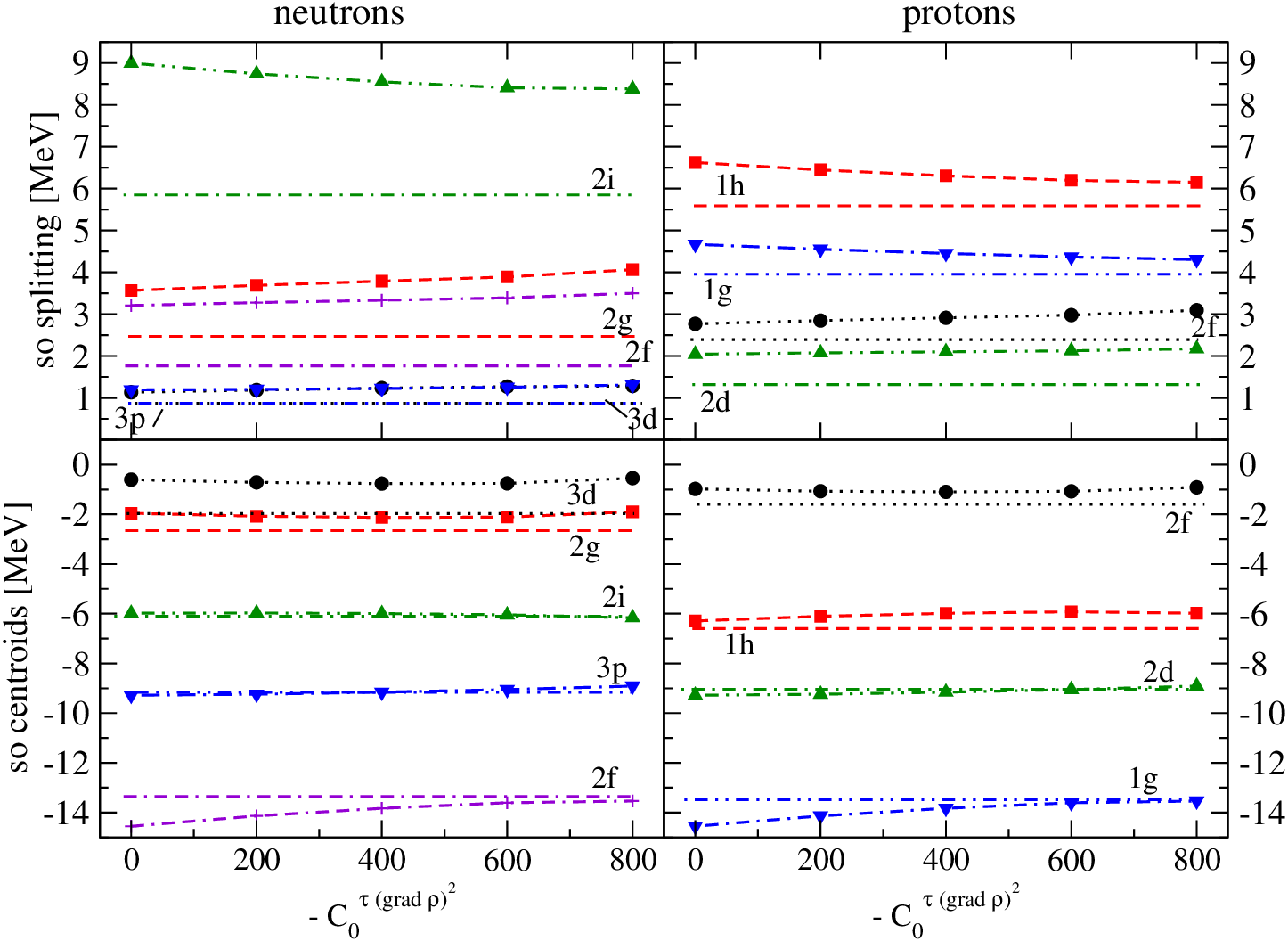}
\end{center}
\caption{Spin-orbit splitting and centroids for $^{208}$Pb as a function of the coefficient
$C_0^{\tau (\nabla \rho)^2}$. The experimental values are taken from \cite{vautherin1972}.}
\label{fig:pbso}
\end{figure}

We now compare our results to the recent ones of Zalewski \textit{et al.}~\cite{zalewski2010,zalewski2010_bis}
where correction terms such as (\ref{eq:hcorr}) as well as others have been studied.
The main differences between our approach and that of Refs.~\cite{zalewski2010,zalewski2010_bis}
are: \textit{(i)} the moderating term is not present in Refs.~\cite{zalewski2010,zalewski2010_bis}, 
\textit{(ii)} the effective mass in the bulk of nuclei is close to 0.8$m$ in our case, while in the
Refs.~\cite{zalewski2010,zalewski2010_bis} it assumes the value 1 since they started with SkX-Skyrme
interaction~\cite{brown1998} , \textit{(iii)} we have not refitted the parameters of the interaction 
contrarily to Refs.~\cite{zalewski2010,zalewski2010_bis}; indeed, at variance with our approach, the functional is readjusted in Refs.~\cite{zalewski2010,zalewski2010_bis} such that the condition
\begin{equation}
\int \! d\mathbf{r} \; \frac{\rho_0(r)}{A} \; \frac{m^*(r)}{m} = 1 \; 
\end{equation}
is satisfied.
An important dependence of the spin-orbit splitting and of the centroids as a function of the 
coefficient of the correction term have been observed in Ref.~\cite{zalewski2010}.
In our case, as shown in Figs.~\ref{fig:caso} and \ref{fig:pbso} for both $^{40}$Ca and $^{208}$Pb
nuclei, we do not find such an important effect. 
Only a weak dependence on the coefficient $C_0^{\rho^2 (\nabla \rho)^2}$ of the spin-orbit 
splitting is observed and an almost independence of the spin-orbit centroids.
This contradiction between the two results might 
come from the readjustement
of the Skyrme parameters 
(point \textit{(iii)} mentioned above) 
performed in Ref.~\cite{zalewski2010_bis}.
In the latter article, the spin-orbit interaction is not changed, but the parameters of the Skyrme
interaction are changed, which might induce a modification of the density profile and, therefore, of
the spin-orbit splitting.
In fact, in Ref.~\cite{zalewski2010_bis} another procedure is adopted, the \textit{non-perturbative}
one, and the behavior of the the spin-orbit splitting and of the spin-orbit centroids
is quite different from that shown in Ref.~\cite{zalewski2010}.
The \textit{non-perturbative} prescription of Ref.~\cite{zalewski2010_bis} shows almost no change
of the spin-orbit splittings and centroids with respect to the strength of the surface-peaked
effective mass.
This shows that the spin-orbit splittings and their centroids 
might not be impacted by the presence 
of a surface-peaked effective mass in a direct way, but eventually, indirectly, through the readjustment
procedure of the functional.

Finally, we have studied how the binding energy varies as a function of the correction term~(\ref{eq:hcorr}).
The results are presented in Table~\ref{tab:be}.
The binding energy of $^{40}$Ca and $^{208}$Pb increases as the parameter
$|C_0^{\tau (\nabla \rho)^2}|$ increases.
We can therefore expect that the readjustment of the parameters of the Skyrme 
interaction shall essentially make 
the interaction
%
more attractive.

\section{\label{sec4} Pairing properties}

Most of the nuclei are superfluid and it could be shown that in the weak coupling limit
of the BCS approximation, the pairing gap at the Fermi surface $\Delta_F$ and the
pairing interaction $v_{pair}$ are related in uniform matter through the 
relation~\cite{lombardo99}:
\begin{equation}
\Delta_F \approx 2 \epsilon_F \exp [2/(N_0 v_{pair})],
\end{equation}
where $\epsilon_F$ is the Fermi energy, $N_0=m^* k_F/(\hbar^2 \pi^2)$ is the density of states
at the Fermi surface.
Then a small change of the effective mass $m^*$ can result in a substantial change of the
pairing gap.

In the following, we will consider $^{120}$Sn because it is an excellent candidate
to study pairing correlations~\cite{bender2003}: $^{120}$Sn is spherical and only neutrons
are participating to the S-wave Cooper-pairs.
An accurate description of the pairing properties can be obtained already at the
level of the spherical HF+BCS framework~\cite{bender2003}.
In this section, we study qualitatively the relation between the increase of the 
effective mass at the surface and its consequences on the pairing properties, both at zero and at finite temperature.

We adopt a density-dependent contact interaction $v_{nn}$ given by~\cite{garrido1999}:
\begin{equation}
\left\langle{k} | v_{nn} | {k'} \right\rangle = v_0 g(\rho) \theta(k,k') \ ,
\label{eq:vpair}
\end{equation}
where the strength $v_0$ of the pairing interaction is adjusted to obtain an average
pairing gap equals to 1.3~MeV in $^{120}$Sn. 
The factor $g(\rho)$ in Eq.~(\ref{eq:vpair}) is a density-dependent function (see below) and $\theta(k,k')$ is the cutoff 
introduced to regularize the ultraviolet divergence in the gap equation.
We choose for $\theta(k,k')$ the following prescription: $\theta(k,k') = 1$ if $E_k,E_{k'}<E_c$, otherwise
it is smoothed out with the Gaussian function $\exp \left(-\left[(E_\lambda-E_c)/a \right]^2 \right)$.
Hereafter, we choose $E_c = 8$~MeV and $a = 1$~MeV.
Notice that to be compatible with HFB calculations, the cutoff is implemented on the quasiparticle energy,
$E_{\lambda,q} = [(\epsilon_{\lambda,q} - \mu_q)^2 + \Delta_{\lambda,q}^2]^{1/2}$, where $\epsilon_\lambda$ is the
HF energy, $\mu$ the chemical potential, and $\Delta_{\lambda,q}$ the average pairing gap for the state
$\lambda$ (see Eq.~(\ref{eq:deltalambdaq})).
The density-dependent term $g(\rho)$ is simply defined as:
\begin{equation}
g(\rho) = 1 - \eta \frac{\rho}{\rho_0} \ ,
\end{equation}
where $\eta$ designates the volume ($\eta = 0$), mixed ($\eta = 0.5$), or surface ($\eta = 1$) 
character of the interaction and $\rho_0 = 0.16\ \mathrm{fm^{-3}}$ is the
saturation density of symmetric nuclear matter.
The isoscalar particle density, $\rho=\rho_n + \rho_p$, is defined as:
\begin{equation}
\rho_q(r) = \frac{1}{4 \pi} \sum_\lambda (2 j_\lambda+1) \left[ v^2_{\lambda,q} (1 - f_{\lambda,q}) + u^2_{\lambda,q} f_{\lambda,q} \right] \vert \phi_{\lambda,q}(r)\vert^2 \; ,
\label{eq:particle}
\end{equation}
where the $u_{\lambda,q}$ and the $v_{\lambda,q}$ are variational parameters.
The $v^2_{\lambda,q}$ represent the probabilities that a pairing state is occupied in a state ($\lambda,q$), $u^2_{\lambda,q}=1-v^2_{\lambda,q}$, and $f_{\lambda,q}$ is the Fermi function for the quasiparticle energy $E_{\lambda,q}$~\cite{goodman1981}:
\begin{equation}
f_{\lambda,q} = \frac{1}{1 + e^{E_{\lambda,q}/{k_B T}}} \ ,
\label{eq:fi}
\end{equation}
where $k_B$ is the Boltzmann constant.\\
The local pairing field is then given by:
\begin{equation}
\Delta_q(r) = \frac{v_0}{2}\ g(\rho) \tilde{\rho}_q(r) \ ,
\label{eq:gapeq}
\end{equation}
where $\tilde{\rho}_q$ is the abnormal density defined as:
\begin{equation}
\tilde{\rho}_q(r) = - \frac{1}{4 \pi} \sum_\lambda (2 j_\lambda+1) u_{\lambda,q} v_{\lambda,q} (1 - 2 f_{\lambda,q}) \vert \phi_{\lambda,q}(r)\vert^2 \ .
\label{rhotilde}
\end{equation}
The 
Eq.~(\ref{eq:gapeq}) is the self-consistent gap equation that should be solved consistently with the 
particle conservation equation~(\ref{eq:particle}).
The average pairing gap for the state $\lambda$ used in the definition of the cutoff is defined as:
\begin{equation}
\Delta_{\lambda,q}=\int \! d\mathbf{r} \vert \phi_{\lambda,q}(r)\vert^2 \Delta_q(r).
\label{eq:deltalambdaq}
\end{equation}

In the HF+BCS framework, Eqs.~(\ref{eq:particle}) and (\ref{eq:gapeq}) are solved at each iteration. 
The number of iterations performed depends on the convergence of the pairing gap 
equation (\ref{eq:gapeq}); it goes from about 200 up to about 1000 near the critical temperature.

\subsection{Pairing properties at $T=0$}

In the following, we study the influence of the correction term~(\ref{eq:hcorr}) on the pairing properties
at $T=0$.
A realistic calculation for nuclei shall treat consistently the pairing interaction in 
the particle-particle channel and that in the particle-hole channel.
The bare interaction in the particle-particle channel shall then be replaced by the induced 
one which accounts for a 50\% correction~\cite{gori2005}.
It is however not our intention in the present work to investigate this question.
We want, at a simpler level, to clarify the role of the correction term~(\ref{eq:hcorr}) on
the pairing properties, and we will show that there is indeed a correlation in space between the
enhancement of the effective mass and that of the probability distribution of the Cooper pairs.

In order to reproduce the value of the average gap $\tilde{\Delta}_n=1.3$~MeV in ${}^{120}$Sn, 
we have adjusted Eq.~(\ref{eq:gapeq}) for three kinds of pairing interactions (volume, mixed, and 
surface), for the functional without the correction term~(\ref{eq:hcorr}).
We obtain for the values $(\eta, v_0)$: (0; -259~MeV fm$^3$), (0.5; -391~MeV fm$^3$), (1; -800~MeV fm$^3$).
In Table~\ref{tab:gap} we report the values for the average neutron pairing gap $\tilde{\Delta}_n$, 
calculated as the average gap over the abnormal density,
\begin{equation}
\tilde{\Delta}_n \equiv \frac{1}{\tilde{N}}\int \! d\mathbf{r} \; \tilde{\rho}_n(\mathbf{r}) \Delta_n(\mathbf{r}) \ ,
\label{deltatilde}
\end{equation}
where $\tilde{N}=\int \! d\mathbf{r} \; \tilde{\rho}_n(\mathbf{r})$,
and for several values of the coefficient $C_0^{\tau (\nabla \rho)^2}$.
In Table~\ref{tab:gap}, it is shown that the effect of the correction term~(\ref{eq:hcorr}) on the
pairing gap is non-negligible. 
The pairing gap is increased by 200 to 700~keV as the coefficient $|C_0^{\tau (\nabla \rho)^2}|$
gets larger. 
An 
%
important 
dependence with respect to the kind of the pairing interaction (volume, mixed, surface)
is also observed. 
The largest effect of the correction term~(\ref{eq:hcorr}) is obtained for the surface pairing gap.
From the results presented in Table~\ref{tab:gap} we notice that the correction term~(\ref{eq:hcorr})
in the functional has an important influence on the average pairing gap.

\begin{figure}[htb]
\begin{center}
\includegraphics[width=12cm]{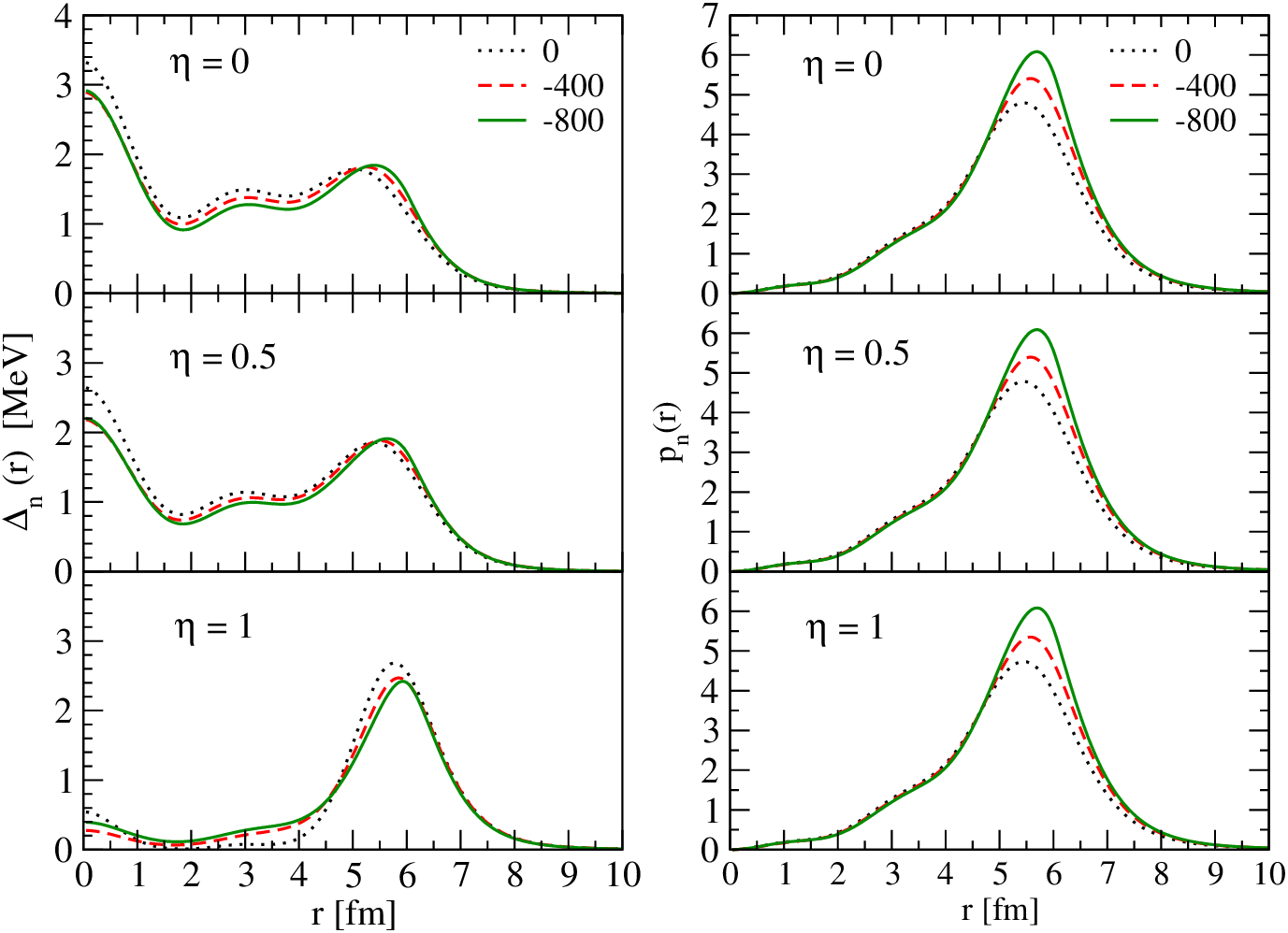}
\end{center}
\caption{Neutron pairing gap (on the left) and probability distribution of the Cooper 
pairs (on the right) for $^{120}$Sn as a function of radial coordinate, for 
$C_0^{\tau (\nabla \rho)^2} =$ 0, -400, -800~MeV fm$^{10}$, and for different 
types of pairing interaction (volume, mixed, surface).}
\label{fig:deltaprob}
\end{figure}

In order to study the effect of the correction term on the pairing properties in a more realistic case, 
we have chosen to renormalize the pairing interaction, for each value of the coefficient 
$C_0^{\tau (\nabla \rho)^2}$, in such a way to get always at zero temperature $\tilde{\Delta}_n=1.3$~MeV.
We obtain, at $T=0$, for $C_0^{\tau (\nabla \rho)^2} = -400 $~MeV fm$^{10}$, the values $(\eta, v_0)$: 
(0; -248~MeV fm$^3$), (0.5; -362~MeV fm$^3$), (1; -670~MeV fm$^3$), and, for 
$C_0^{\tau (\nabla \rho)^2} = -800 $~MeV fm$^{10}$, the values $(\eta, v_0)$: (0; -235~MeV fm$^3$), 
(0.5; -337~MeV fm$^3$), (1; -593~MeV fm$^3$).
We represent in Fig.~\ref{fig:deltaprob} the pairing field and the probability distribution 
of Cooper-pairs defined as
\begin{equation}
p(r) = - 4\pi r^2 \tilde{\rho}(r) \; ,
\label{eq:probab}
\end{equation}
versus the radial coordinate for different values of the parameter $C_0^{\tau (\nabla \rho)^2}$.
On the left panels, we display the pairing field profiles, which do not change very much with the 
coefficient $C_0^{\tau (\nabla \rho)^2}$; this is due to our renormalization procedure, which 
requires the average pairing gap to be 1.3~MeV.
The right panels of Fig.~\ref{fig:deltaprob} clearly show the correlation in space between the 
enhancement of the effective mass and that of the probability distribution of the Cooper pairs, 
i.e. the enhancement of the probability distribution is located where the effective mass is surface-peaked.

In conclusion, it has been shown in this section that the effect of the surface-peaked 
effective mass on the pairing properties is non-negligible. 
As for approaches where the pairing interaction is empirically adjusted on some nuclei, 
we absorbed this effect through a renormalization of the pairing interaction.
For non empirical approaches where the pairing interaction is not adjusted on nuclei 
properties but directly to that of the bare $^1$S$_0$
potential~\cite{margueron2007,margueron2008,lesinski2009,hebeler2009,bertulani2009}, the
enhancement of the average pairing gap induced by the surface-peaked effective mass shall
be treated consistently with the induced pairing interaction~\cite{gori2005}.

\subsection{\label{sec5} Pairing properties at finite temperature}

\begin{figure}[htb]
\begin{center}
\includegraphics[width=12cm]{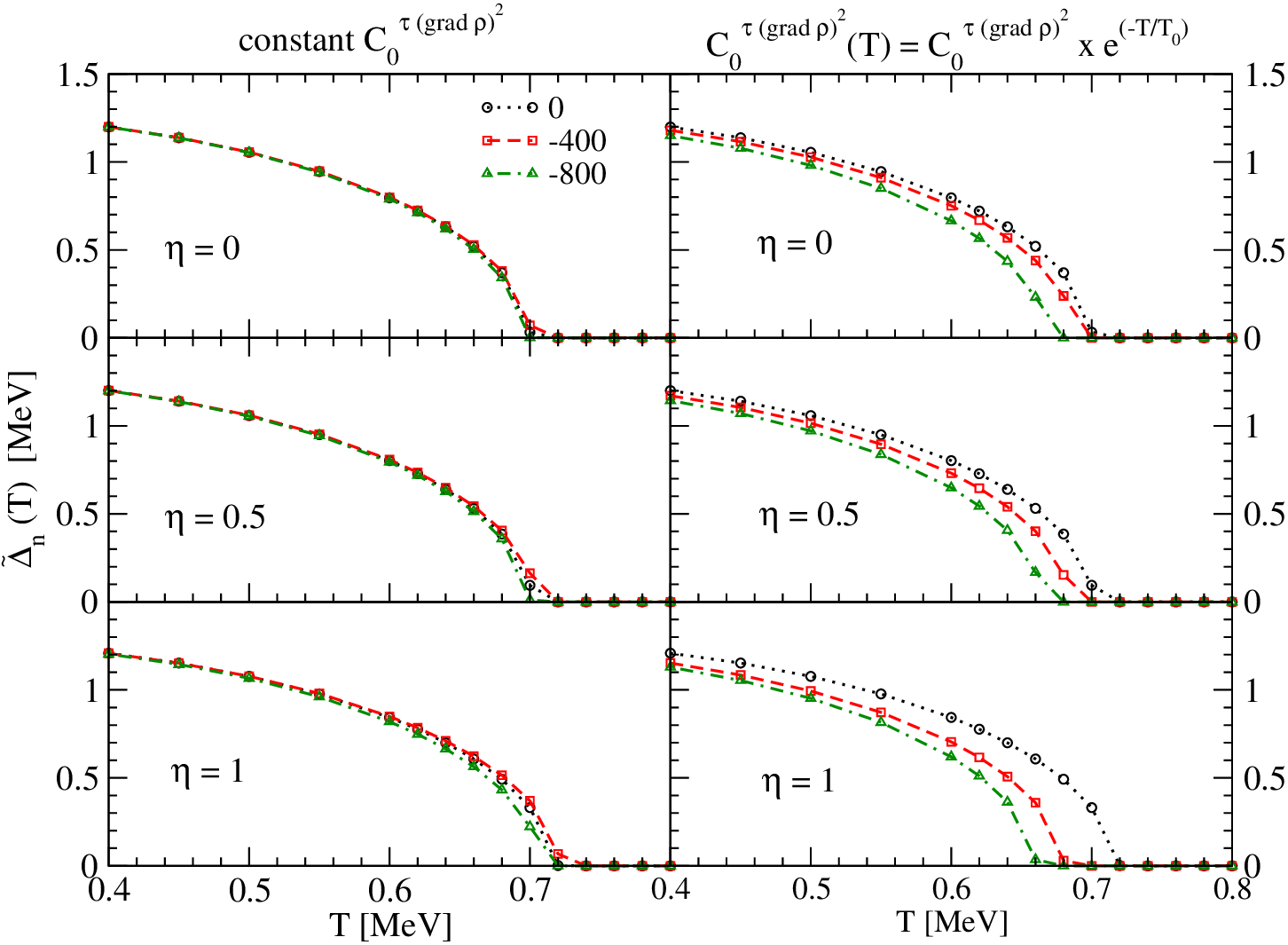}
\end{center}
\caption{Neutron pairing gap for $^{120}$Sn as a function of temperature for different 
values of the coefficient $C_0^{\tau (\nabla \rho)^2}$. The results on the left are obtained 
keeping $C_0^{\tau (\nabla \rho)^2}$ constant ($C_0^{\tau (\nabla \rho)^2}$ = 0, -400, 
-800~MeV fm$^{10}$), while the results on the right are obtained using for 
$C_0^{\tau (\nabla \rho)^2}$ the prescription in Eq.~(\ref{eq:c0T}).}
\label{fig:gapsnt}
\end{figure}

The density of states around the Fermi energy shall influence the temperature-related 
quantities such as the entropy and the specific heat~\cite{chamel2009}, and at the same 
time the density of states is also affected by the temperature~\cite{vinhmau1985,donati1994,giovanardi1996}.
%
%
We explore the effect of the new term on $^{120}$Sn in the framework of 
HF+BCS at finite temperature, using the pairing interaction where the strength has been 
renormalized at $T=0$ for each value of the coefficient $C_0^{\tau (\nabla \rho)^2}$.
In Fig.~\ref{fig:gapsnt} we show the neutron pairing gap as a function of temperature, for the 
three different kinds of pairing interactions (volume, mixed, surface). 
The results on the left are obtained keeping the coefficient $C_0^{\tau (\nabla \rho)^2}$ constant, 
while, on the right, we plot the results obtained letting $C_0^{\tau (\nabla \rho)^2}$ vary 
exponentially with temperature, according to the following relation:
\begin{equation}
C_0^{\tau (\nabla \rho)^2}(T) = C_0^{\tau (\nabla \rho)^2} \times e^{-T/T_0} \ ,
\label{eq:c0T}
\end{equation}
with $T_0=2$~MeV.
The choice of this kind of temperature dependence relies on the work by Donati 
\textit{et al.}~\cite{donati1994}, where a study of the $\omega$-mass in the framework of QRPA 
for temperatures up to some MeV was carried out on some neutron rich nuclei.
The variation of $m_\omega$ with respect to temperature was parameterized with an exponential 
profile and the typical scale of the variation of $m_\omega$ with temperature was found to 
be around 2~MeV.
Notice that in semi-infinite nuclear matter, the typical scale was found to be 
1.1~MeV~\cite{giovanardi1996}. 
The reduction  of the scale might be due to the semi-infinite model which is still far
from a realistic finite nucleus case.

We observe in all cases in Fig.~\ref{fig:gapsnt} the typical behavior associated to the 
existence of a critical temperature $T_c$ after which pairing correlations are 
destroyed~\cite{goodman1981}.
In particular, as expected, $T_c$ is not modified by the new term for the case of constant 
coefficient, since we absorbed the effect of the new term through the renormalization procedure.
Instead, the effect of a temperature dependent coefficient $C_0^{\tau (\nabla \rho)^2}(T)$ is to 
reduce the critical temperature; indeed, the chosen dependence (\ref{eq:c0T}) shifts the critical 
temperature of $\sim$ 40~keV in the case of volume interaction and of $\sim $ 60~keV in 
the case of surface interaction.
The relation $T_c \simeq \tilde{\Delta}_n(T=0)/2$ is still verified; more precisely, the ratio 
$T_c/\tilde{\Delta}_n(T=0)$ is $\simeq 0.55$ (for volume interaction) and $\simeq 0.57$ 
(for surface interaction) for the case of $T$-independent $C_0^{\tau (\nabla \rho)^2}$, 
while with the $T$-dependent prescription~(\ref{eq:c0T}) it varies from 
$\simeq 0.55$ (for volume interaction) to $\simeq 0.51$ (for surface interaction).
In conclusion, we remark that through its temperature dependence, the surface-peaked effective
mass has an effect on the critical temperature, that could also be extracted 
experimentally~\cite{guttormsen2001,schiller2001}.
This perspective motivates the application of the present work to realistic cases
at finite temperature.

\begin{figure}[htb]
\begin{center}
\includegraphics[width=12cm]{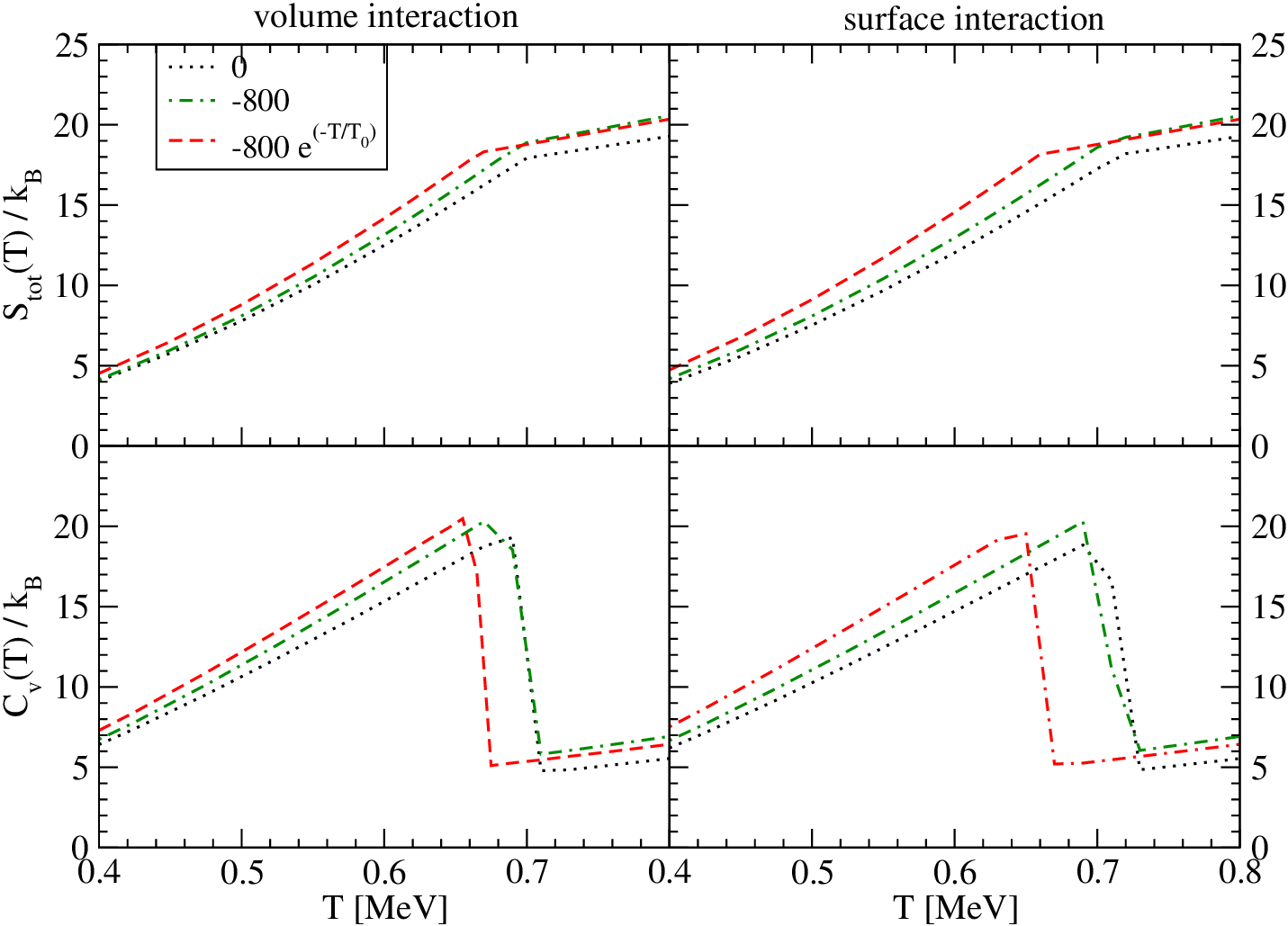}
\end{center}
\caption{Total entropy $S_{tot}$ and specific heat $C_V$ (in units of the Boltzmann constant),
for $^{120}$Sn as a function of temperature, for $C_0^{\tau (\nabla \rho)^2} =$ 0, 
-800~MeV fm$^{10}$, -800~$e^{-T/T_0}$~MeV fm$^{10}$, and for volume (on the left) and 
surface (on the right) pairing interaction.}
\label{fig:entrcvsn}
\end{figure}

Finally, in Fig.~\ref{fig:entrcvsn} we show the total entropy and specific heat as a 
function of temperature for volume and surface interaction, and for three values of 
$C_0^{\tau (\nabla \rho)^2}$ (0, -800~MeV fm$^{10}$, and $C_0^{\tau (\nabla \rho)^2}(T)$ 
for the case -800~MeV fm$^{10}$).
The entropy of the system is calculated as: $S_{tot} = S_n + S_p$, being:
\begin{equation}
S_q = -k_B\ \sum_{\lambda} \left[ f_{\lambda,q} \ln(f_{\lambda,q}) + (1-f_{\lambda,q}) \ln(1 - f_{\lambda,q}) \right] \ ,
\label{eq:entropy}
\end{equation}
where $f_{\lambda,q}$ is defined as in Eq.~(\ref{eq:fi}).
The specific heat is then defined as:
\begin{equation}
C_V = T\ \frac{\partial S_{tot}}{\partial T} \ .
\label{eq:cv}
\end{equation}
We observe the change of the slope in the entropy curve, which causes the discontinuity in the 
specific heat in correspondence of the critical temperature.
In agreement with the previous results, $T_c$ is shown to be modified by a 
temperature-dependent $C_0^{\tau (\nabla \rho)^2}$, the effect is stronger in the case of surface 
interaction, and the temperature dependent coefficient acts as to reduce $T_c$.

\section{\label{sec6} Conclusions}

In this paper we have studied the influence of the correction term~(\ref{eq:hcorr}) on various nuclear
properties.
The isoscalar correction term~(\ref{eq:hcorr}) has been shown to produce a surface-peaked effective 
mass in the nuclei under study ($^{40}$Ca and $^{208}$Pb), without modifying significantly 
the mean field profiles.
The increase of $m^*/m$ at the surface is up to about 1.2 - 1.3 for the maximum value of the strength of the 
correction we used.
As the effective mass gets enhanced at the surface of nuclei, the density of states increases.
Then, we have studied the impact of such a term on the neutron pairing gap in the semi-magic nucleus 
$^{120}$Sn, within an HF+BCS framework, and it turned out that its effect is non-negligible; if the 
pairing interaction is not renormalized consistently with the new term, the average gap increases 
from 200~keV to 700~keV under variation of the strength of the correction term and depending on 
the volume/surface character of the interaction.
In a recent work~\cite{baldo2010}, it has been stressed that the surface enhancement of 
the pairing field induced by the PVC might also play a role on the size of the Cooper-pairs at the surface
of nuclei. 
In uniform matter, the coherence length is indeed inversely proportional to the pairing gap.
The surface enhancement of the pairing field could then make the Cooper-pairs smaller at the surface
of nuclei.
It would be interesting to investigate whether this interesting feature of the pairing correlation might be probed experimentally, for example by pair transfer reaction mechanism.
%
Finally, we have explored some finite temperature properties in $^{120}$Sn, within a HF+BCS 
framework at finite temperature.
We observed for the neutron pairing gap that the critical temperature at which pairing correlations 
vanish is shifted if a $T$-dependence in the new coefficient is considered.
As a consequence, the entropy and specific heat profiles are affected by the introduction of the new term.

In the future, we shall go on with a global refitting of all the parameters of the functional, including the new
correction term~(\ref{eq:hcorr}). 
It would be very instructive to know whether both the masses and the single particle levels could 
be improved in such a way.

A. F. F. and J. M. would like to thank E. Khan and N. Sandulescu for their help in
checking our HF+BCS results, as well as the fruitful discussions with N. Van Giai and M. Grasso.
This work was supported by CompStar, a Research Networking Programm of the European Science Foundation,
by the ANR NExEN, and by the exchanged fellowship program of the Universit\'e Paris-Sud XI.

\newpage

\begin{table}[htb]
\caption{Charge rms radius, $r_\mathrm{ch}$, and neutron skin radius, $r_\mathrm{skin}$, 
for $^{40}$Ca and $^{208}$Pb and for different values of the coefficient 
$C_0^{\tau (\nabla \rho)^2}$.
The charge rms radius is calculated according to Eq.~(110) in Ref.~\cite{bender2003}. \label{tab:rms}}
\setlength{\tabcolsep}{.3in}
\begin{tabular}{ccccc}
\hline
& \multicolumn{2}{c}{$^{40}$Ca} & \multicolumn{2}{c}{$^{208}$Pb} \\
$C_0^{\tau (\nabla \rho)^2}$ & $r_\mathrm{ch}$ & $r_\mathrm{skin}$ & $r_\mathrm{ch}$ & $r_\mathrm{skin}$ \\
$[\mathrm{MeV fm}^{10}]$ & [fm] & [fm] & [fm] & [fm] \\  
\hline
0	& 4.01 & -0.04 & 6.05 & 0.16 \\
-200	& 4.05 & -0.04 & 6.08 & 0.16 \\
-400	& 4.08 & -0.04 & 6.10 & 0.16 \\
-600	& 4.10 & -0.04 & 6.12 & 0.16 \\
-800	& 4.10 & -0.04 & 6.12 & 0.16 \\
\hline
\end{tabular}
\end{table}

\begin{table}[htb]
\caption{Binding energy per nucleon in $^{40}$Ca and $^{208}$Pb (in MeV) for different 
values of the coefficient $C_0^{\tau (\nabla \rho)^2}$.\label{tab:be}}
\setlength{\tabcolsep}{.3in}
\begin{tabular}{ccc}
\hline
$C_0^{\tau (\nabla \rho)^2}$ &  $^{40}$Ca & $^{208}$Pb \\
\hline
0	 & -8.781 & -8.071 \\
-200 & -8.591 & -7.983 \\
-400 & -8.442 & -7.910 \\
-600 & -8.322 & -7.849 \\
-800 & -8.227 & -7.801 \\
\hline
\end{tabular}
\end{table}

\begin{table}[htb]
\caption{Average neutron pairing gap $\tilde{\Delta}_n$ for ${}^{120}$Sn for different values 
of the coefficient $C_0^{\tau (\nabla \rho)^2}$. \label{tab:gap}}
\setlength{\tabcolsep}{.3in}
\begin{tabular}{cccc}
\hline
$C_0^{\tau (\nabla \rho)^2}$ & \multicolumn{3}{c}{$\eta$} \\
~$[$MeV fm$^{10}]$ & 0 & 0.5 & 1 \\
\hline
0	& 1.30 & 1.30 & 1.30 \\
-200	& 1.33 & 1.36 & 1.47 \\
-400	& 1.37 & 1.43 & 1.62 \\
-600	& 1.42 & 1.52 & 1.79 \\
-800	& 1.49 & 1.60 & 1.96 \\
\hline
\end{tabular}
\end{table}

\end{document}